\documentclass[%
superscriptaddress,
showpacs,preprintnumbers,
amsmath,amssymb,
aps,
prl,
twocolumn
]{revtex4-1}
\usepackage{ulem}
\usepackage{graphicx}% Include figure files
\usepackage{dcolumn}% Align table columns on decimal point
\usepackage{bm}% bold math
\usepackage{color}
\begin {document}
% \section{title}
%\preprint{APS/123-QED}

\title{
Thermal transport of confined water molecules in quasi-one-dimensional nanotubes
}% Force line breaks with \\
%\thanks{A footnote to the article title}%

\author{Shun Imamura}
\affiliation{Department of System Design Engineering, Keio University, Yokohama, Kanagawa 223-8522, Japan}
\author{Yusei Kobayashi}
\affiliation{Faculty of Mechanical Engineering, Kyoto Institute of Technology, Sakyo-ku, Kyoto, 606-8585, Japan}
\author{Eiji Yamamoto}
\email{eiji.yamamoto@sd.keio.ac.jp}
\affiliation{Department of System Design Engineering, Keio University, Yokohama, Kanagawa 223-8522, Japan}
% \affiliation{
%  Department of System Design Engineering, Keio University, Yokohama, Kanagawa 223-8522, Japan
% }

%\collaboration{MUSO Collaboration}%\noaffiliation

%\date{\today}% It is always \today, today,
%  but any date may be explicitly specified

\begin{abstract}
Dimensions and molecular structure play pivotal roles in the principle of heat conduction. The dimensional characteristics of solution within nanoscale systems depend on the degrees of confinement. However, the influence of such variations on heat transfer remains inadequately understood. Here, we perform quasi-one-dimensional non-equilibrium molecular dynamics simulations to calculate the thermal conductivity of water molecules confined in carbon nanotubes. The structure of water molecules is determined depending on the nanotube radius, forming a single-file, a single-layer, and a double-layer structure, corresponding to an increasing radius order. We reveal that the thermal conductivity of liquid water has a sublinear dependency on nanotube length exclusively when water molecules form a single file. Stronger confinement leads to behavioral and structural characteristics closely resembling a one-dimensional nature. Moreover, single-layer-structured water molecules exhibit enhanced thermal conductivity. We elucidate that this is due to the increase in the local water density and the absence of transitions to another layer, which typically occurs in systems with double-layer water structures within relatively large radius nanotubes.
\end{abstract}

% %\pacs{87.16.dj,87.15.ap,68.35.Fx,02.50.-r}
% %02.50.-r		Probability theory, stochastic processes, and statistics (see also section 05 Statistical physics, thermodynamics, and nonlinear dynamical systems)
% %05.40.-a 	Fluctuation phenomena, random processes, noise, and Brownian motion (for fluctuations in superconductivity, see 74.40.-n		for statistical theory and fluctuations in nuclear reactions, see 24.60.-k; for fluctuations in plasma, see 52.25.Gj; for nonlinear dynamics and chaos, see 05.45.-a)
% %05.45.-a 	Nonlinear dynamics and chaos (see also section 45 Classical mechanics of discrete systems; for chaos in fluid dynamics, see 47.52.+j; for chaos in superconductivity, see 74.40.De)
% %05.60.-k 	Transport processes
% %68.35.Fx 	Diffusion; interface formation (see also 66.30.-h Diffusion in solids, for diffusion of adsorbates, see 68.43.Jk)
% %87.10.Tf     	Molecular dynamics simulation
% %87.10.Mn 	Stochastic modeling
% %87.14.Cc 	Lipids
% %87.14.E- 	Proteins
% %87.14.ep 	Membrane proteins
% %87.15.A- 	Theory, modeling, and computer simulation
% %87.15.ap 	Molecular dynamics simulation (biological)
% %87.15.H- 	Dynamics of biomolecules
% %87.15.hj 	Transport dynamics
% %87.15.K- 	Molecular interactions; membrane-protein interactions
% %87.15.kt 	Protein-membrane interactions
% %87.15.Vv 	Diffusion
% %87.16.dj 	Dynamics and fluctuations
% % Classification Scheme.
% \keywords{Suggested keywords}%Use showkeys class option if keyword
% %display desired
\maketitle

\section{Introduction}
Fourier's law describes the conduction of heat in a substance with a temperature gradient. The thermal energy proportional to the temperature gradient flows from higher to lower temperature, which is expressed as
\begin{equation}
    J=-\kappa \nabla T,
\end{equation}
where $J$ is the heat current, $\nabla T$ is the temperature gradient, and $\kappa$ is the thermal conductivity. It has been proved that $\kappa$ takes a constant value independent of system size in a three-dimensional (3D) bulk material. 

In low dimensional systems, however, Fourier's law does not hold and $\kappa$ diverges with the system length $L$ as a power law: $\kappa \sim L^{\alpha}$ \cite{lepri2003thermal, li2003anomalous, dhar2008heat, lepri1997heat, lepri2016thermal}. In a 1D system exhibiting anomalous heat transport, the value of $\alpha$ is related to the Kardar-Parisi-Zhang (KPZ) class \cite{van2012exact, quastel2015one, li2020entropic}. Although the KPZ equation originally describes interfacial growth, it appears in many models of non-equilibrium systems, such as directed polymers, fluids with fluctuations, and 1D heat transport phenomena \cite{gueudre2012directed, mendl2013dynamic, van2012exact}. The KPZ scaling law is classified into two classes, and different values of $\alpha =1/3$ and $\alpha =1/2$ are theoretically derived for each class. The values of $\alpha$ actually obtained in simulations and experiments are within the range of 0.3 to 0.5, even if they are not exactly these theoretical values. Such anomalous heat transport has been actually demonstrated in numerical simulations using 1D chain models \cite{wang2004intriguing}, isolated polymer chains \cite{henry2008high}, and quasi-1D carbon nanotubes (CNT) \cite{sevik2011phonon, meng2021thermal}. It has also been observed in many experiments on semiconductor nanowires \cite{hsiao2013observation}, polymer nanofibers \cite{tang2016length}, and CNT \cite{lee2017divergent}. Furthermore, a recent study has discovered that anomalous heat conduction occurs even in elongated nanoscale 3D lattice and fluid systems closed to 1D systems with the multiparticle collision dynamics simulation \cite{wang2010heat, lepri2021kinetic, luo2021heat}. In particular, the results in 3D fluid systems could be important in considering the mechanism of thermal conduction, but the simulation used simple particles that only apply collisions as interactions between particles, so the behavior might differ from the actual molecules. 

Solutions in nanoscale confined systems have different properties from bulk systems. Since the ratio of the surface area to the volume of fluids is unusually high, surface characteristics, including wettability and roughness have a critical influence on the transport of matter and energy \cite{sun2020nanoconfined, mohammad2020anomalous, fu2018understanding, sun2012transport, wu2017wettability}. In addition, the liquid molecules in the confined system form peculiar structures depending on the conditions, such as the size and shape of the system, pressure, and external electric field. For example, molecular dynamics (MD) simulations have shown that solution molecules in CNT form helical and ice-like regular structures \cite{liu2005fluid, winarto2015structures, takaiwa2015water, WinartoYamamotoYasuoka2017}. In particular, water molecules exhibit remarkably unique thermophysical properties in nanoconfined systems because of their polarity and anisotropy \cite{hu2010water, zhao2020hierarchical, zhao2020thermal}. However, there is limited understanding regarding the impact of these characteristics on heat conduction, encompassing anomalous heat transport in solutions.

\begin{figure*}[t]
    \centering
    \includegraphics[width=0.9\textwidth,bb= 0 0 615 412]{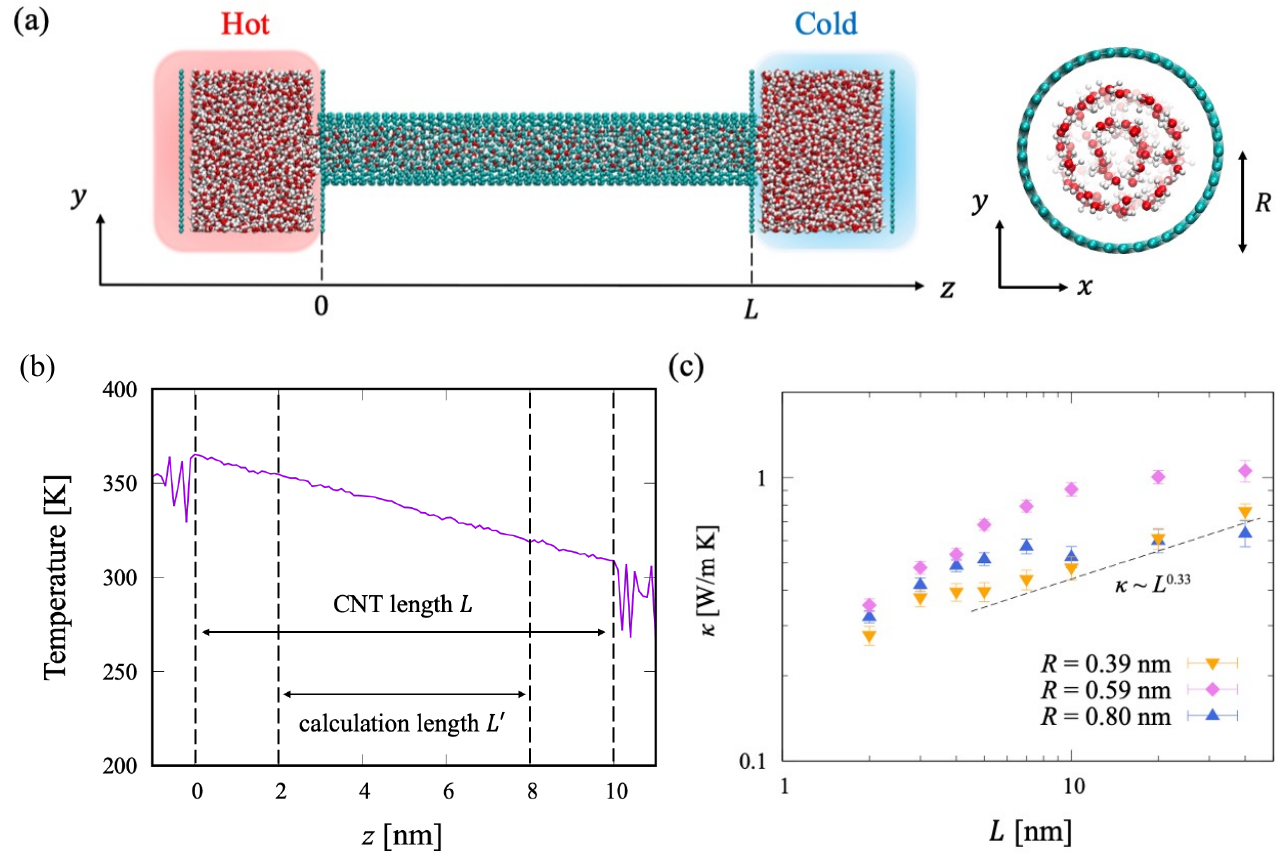}
    \caption{\label{fig:wide} Thermal conductivity dependence on CNT length. (a) Illustration of a simulation system. The CNT has a length of $L=10 \, \mathrm{nm}$ and a radius of $R=0.80 \, \mathrm{nm}$ in this figure. Periodic boundary conditions are applied in the $x$ and $y$ directions, while free boundary conditions is applied in the $z$ direction. An inward pressure $P_{zz}=1.0 \, \mathrm{MPa}$ is applied by each piston. (b) Temperature profiles of water in CNT of $L=10 \, \mathrm{nm}$ and $R=0.80 \, \mathrm{nm}$. As carbon atoms are fixed, they have no temperature. (c) The thermal conductivity in the $z$ direction with different $L$. The dashed line indicates $\kappa \sim L^{0.33}$, which is the theoretical value based on the KPZ class. Error bars are calculated as the standard deviation of the thermal conductivity computed every $1 \, \mathrm{ns}$ for $30 \, \mathrm{ns}$ of collected data.} 
    \label{fig:fig1}
\end{figure*}

In this work, we explore the heat conduction of liquid water in quasi-1D systems, with various degrees of confinement leading to differences in the number of water molecules aligned side by side. We perform non-equilibrium MD simulations with a temperature gradient applied to the water molecules inside the carbon nanotubes in order to investigate the effects of the radius and length of CNT on the heat transport of water. Our results show that thermal conductivity increases with CNT length only in the case that water molecules form a single-file structure, located one-dimensionally like a single chain. In near-one-dimensional systems, the thermal conductivity of water is proportional to $L^{\alpha}$ even in nanoconfined systems, similar to the one-dimensional systems presented so far, and the value of $\alpha = 0.33$ is consistent with the theory based on the KPZ class. Furthermore, we reveal that the thermal conductivity is remarkably higher in the single-layer structure than in any other structure. Water molecules form different structures depending on the radius of CNT, which affects the thermal conductivity.

\section{METHODS}
\subsection{MD simulation}
Graphene membranes were attached on both sides of the single-walled zigzag CNT, and a piston was placed outside each reservoir containing water as shown in Fig.~\ref{fig:fig1}a. The carbon atoms composing CNT and graphene membranes were fixed in space and time to consider the heat transport of liquid water in insulating materials. The pistons move only in the $z$ direction, and the bonding state of the constituent carbon atoms remains unchanged throughout the process. In the $x$ and $y$ directions, periodic boundary conditions are applied, while free boundary conditions in the $z$ direction. In order to adapt the graphene sheet to the periodic boundary, the lateral lengths $l_x$ and $l_y$ were different, $l_{x}=3.829 \, \mathrm{nm}$ and $l_{y}=3.684 \, \mathrm{nm}$ for all systems. 

 All simulation systems composed of carbon nanotube, graphene sheets, and water molecules were established by using VMD build-in tools \cite{humphrey1996vmd}. All simulations were performed with the large-scale atomic/molecular massively parallel simulator (LAMMPS) \cite{plimpton1995fast}. Water-carbon interactions are described by Lennard-Jones (L-J) 6-12 potential: $\epsilon_{CO}=0.4843 \, \mathrm{kJ\, mol^{-1}}$ and $\sigma_{CO}=0.3283 \, \mathrm{nm}$ \cite{koga2002does}. Water-water interaction is modeled by L-J potential with long-range Coulombic interaction \cite{koga2002does, van1998systematic}. We employed the SPC/E water model \cite{berendsen1987missing}, generally used in thermal simulations, and the SHAKE algorithm \cite{ryckaert1977numerical} was employed to fix the length and the angle of bonds between the atoms of the water molecules. We used a cutoff radius of $0.9 \, \mathrm{nm}$ for the L-J interaction and $0.85 \, \mathrm{nm}$ for the real-space Coulomb interaction in systems with  CNT lengths from $L=2$ to $~20 \, \mathrm{nm}$. To generate an accurate temperature gradient, we set the cutoff radius of the Coulomb interaction to $1.9 \, \mathrm{nm}$ for a system with the CNT length $L=40 \, \mathrm{nm}$. Long-range Coulomb interaction was treated using the particle-particle particle-mesh (PPPM) method \cite{hockney1989particle}. 
 
Water molecules initially placed in each reservoir were packed into the CNT by pistons during \textit{NPT} simulations for $2 \, \mathrm{ns}$ at equilibrium with a constant entire temperature $T_{eq}=323 \, \mathrm{K}$ using a Nose-Hoover thermostat and $P_{zz}=1.0 \, \mathrm{MPa}$ applied by pistons. This high pressure by the piston was applied during not only equilibrium simulations but also non-equilibrium simulations to prevent the generation of bubbles within the CNT. After making sure that the nanotube had been filled with water, we used a canonical sampling thermostat that uses global velocity rescaling with Hamiltonian dynamics \cite{bussi2007canonical} to control the temperature, setting the water molecules in the left reservoir to $353 \, \mathrm{K}$ and the right reservoir to $293 \, \mathrm{K}$. We ran non-equilibrium MD simulations with a temperature gradient until a steady state from $30 \, \mathrm{ns}$ up to $100 \, \mathrm{ns}$ depending on the radius and length of the CNT. In some systems, water molecules move from one reservoir to the other reservoir when a temperature gradient is applied, and the pistons also move correspondingly. A steady state was defined as the state more than $30 \, \mathrm{ns}$ after the pistons stopped moving (Fig. S1). A list of the times consumed to reach steady state for each system is shown in TABLE S1 and S2. Then, subsequent simulation were performed for $30 \, \mathrm{ns}$ to collect data. The time step was $2 \, \mathrm{fs}$ in all phases.

 \subsection{Calculation of thermal conductivity}
 We calculated $\kappa$, the thermal conductivity of water molecules within the CNT in the $z$ direction, from Eq.(1) with no flow by means of pressure from pistons on both sides. The heat flux in the $z$ direction $J_{z}$ can be expressed as follows \cite{irving1950statistical,kjelstrup2008non}:
\begin{equation}
    J_{z}= \frac{1}{V} \left[ \sum_{i \in V }e_{i}v_{i, z} + \frac{1}{2} \sum_{i<j \in V} \left( \bm{F}_{ij} \cdot (\bm{v}_{i}+\bm{v}_{j}) \right)r_{ij, z} \right], 
\end{equation}
where $e_{i}$, $v_{i, z}$ represent the energy (kinetic and potential), the velocity in the $z$ direction of the particle $i$, respectively. $ \bm{v}_{i}$ and $ \bm{v}_{j}$ are the velocity vector of the particle $i$ and $j$. $ \bm{F}_{ij}$, $r_{ij, z}$ represent the force, the distance in the $z$ direction between $i$-$j$ particles, respectively. $V$ is the volume of the CNT ($=\pi R^2 L'$). Here, $R$ is the radius of CNT and $L'$ is the calculation length of CNT. Considering the instability of temperature in the vicinity of the CNT junction, water molecules around there were excluded from our analysis. Therefore, $L'$ is shorter than the nanotube length $L$ due to this excluded region (see Fig. 1b). The summation is calculated for particles $i$ and,~ $j$ (oxygen and hydrogen atoms) within the CNT. 

Figure 1b shows the temperature profile for water along the $z$-axis. The temperature profile is linear within the CNT, but it fluctuates spatially due to the small number of water molecules in the connecting part. Therefore, we used the temperature gradient $\nabla T$ within calculation length for analyzing the thermal conductivity. In addition, the temperature of the water molecules located at the edge of the CNT is about $10$ K higher than inside each reservoir. This is because the density of water molecules in the nanotube is significantly lower than in the reservoir. Other systems using CNT with different radii and lengths show similar tendencies in temperature profiles.

\section{Results}
Figure 1c presents the CNT length dependence of thermal conductivity in each radius. In a system with radius $R=0.39 \, \mathrm{nm}$, that is, close to a one-dimensional system, $\kappa$ increases sublinearly in large $L$ and does not converge. $\kappa$ is proportional to 0.33 to the power of $L$, and the value $\alpha=0.33$ is consistent with previous theoretical and experimental research of 1D particle chain models and solid systems \cite{meng2021thermal, lee2017divergent, benenti2013conservation}. Besides, $\kappa$ grows with a similar slope for all systems in the range $L<3 \, \mathrm{nm}$. In a very short system, the heat flow behaves like ballistic transport ($\alpha=1$). Its relaxation process appears in these results without dependence on the CNT radius. To the contrary, for $R = 0.59 \, \mathrm{nm}$ and $R=0.80 \, \mathrm{nm}$, $\kappa$ converges to a certain value in large $L$ in the same way as water in the bulk system even within the constraints of the nanoconfined system. For $R=0.59 \, \mathrm{nm}$, interestingly, $\kappa$ reaches a plateau at $L$ larger than $R=0.80 \, \mathrm{nm}$ and the convergence of $\kappa$ is also higher. That implies that the radius of the system influences the behavior of the thermal conductivity of liquid within nanoconfined systems. Here, rather than merely depending on the radius size, it is inferred that the structure formed by the confined liquid affects on the thermal property.

\begin{figure}
    \centering
    \includegraphics[width=0.5\textwidth,,bb= 0 0 281 312]{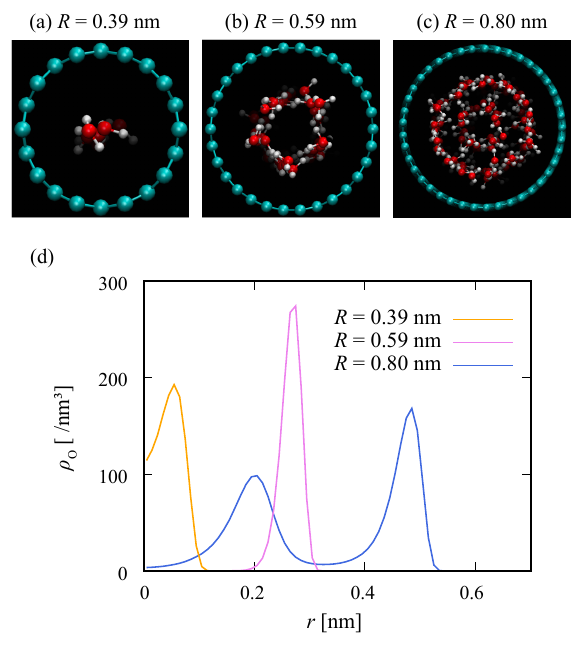}
    \caption{\label{fig:wide}
    Structure of water molecules inside CNTs. The results shown here are for $L=10 \, \mathrm{nm}$. (a-c) Snapshots of cross-sections perpendicular to the $z$-axis for each radius. (d) Number density distribution function of oxygen atoms, where $r$ is the radial distance from the central axis of CNT. From the position and number of peaks observed, it is apparent that water molecules form a single-file, single-layer, and double-layer structure for $R=0.39, 0.59$, and $0.80 \, \mathrm{nm}$, respectively. }
    \label{fig:fig2}
\end{figure}

We investigate how water molecules in the CNT form a structure for each radius $R$. The results for $L=10 \, \mathrm{nm}$ are shown here, but the results are almost the same for other $L$ values. Figures~\ref{fig:fig2}a-c show the typical cross-sectional configuration perpendicular to the CNT.  For $R=0.39 \, \mathrm{nm}$, a single file is formed one-dimensionally like a single chain located in the center of the nanotube. As the radius of the CNT increases, the water molecules get aligned in a circular shape. Specifically, a single-layer and double-layer configuration appear at $R = 0.59 \, \mathrm{nm}$ and $R = 0.80 \, \mathrm{nm}$, respectively. Corresponding to each structure, the number density of oxygen atoms against the distance $r$ from the central axis of CNT has one peak around $r=0.05 \, \mathrm{nm}$ for $R=0.39 \, \mathrm{nm}$ and around $r=0.25 \, \mathrm{nm}$ for $R=0.59 \, \mathrm{nm}$ (Fig. 2d). There are two peaks for $R=0.80 \, \mathrm{nm}$, one peak around $r=0.2 \, \mathrm{nm}$ is lower than the other around $r=0.5 \, \mathrm{nm}$. Such density distributions have been observed in other simulations of similar systems \cite{striolo2006simulated, hanasaki2006flow}, and are independent of the presence or absence of temperature gradients.

\begin{figure}
    \centering
    \includegraphics[width=0.5\textwidth,,bb= 0 0 268 398]{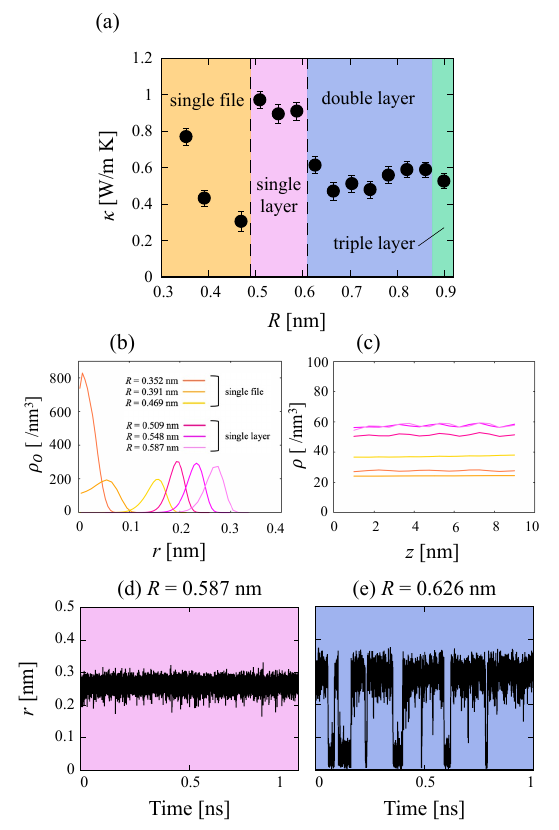}
    \caption{\label{fig:wide}
    Dependencies of the nanotube radius on the thermal conductivity for systems $L=10 \, \mathrm{nm}$. (a) Thermal conductivity with different $R$. The structure of the water molecules varies with $R$ and is divided into three types within the range of our simulations. (b, c) Comparison of the density of water molecules forming a single-file and a single-layer structure. (b) Number density distribution function of oxygen atoms, in the radial direction. (c) Number density distribution function of atoms (oxygen and hydrogen) in the longitudinal direction. The correspondence between line color and radius is the same as in Fig. 3b. (d, e) Distance of an oxygen atom from the central axis of the CNT. Differences in behavior of water molecules forming a single-layer and a double-layer structure.}
    \label{fig:fig3}
\end{figure}

We explore the radius dependence of $\kappa$ in a system with constant $L=10 \, \mathrm{nm}$, the point at which $\kappa$ almost plateaus, to provide a detailed clarification of the influence exerted by the extent of liquid confinement in a quasi-one-dimensional system. As mentioned above, the structure of water molecules is classified into three types (To be precise, at around $R=0.65 \, \mathrm{nm}$, a structure is at the transition point from a single layer to a double layer, where the inner layer of the double layer becomes a single chain. However, since the value of thermal conductivity is close to that of a double-layer structure, both are treated as a double layer in this paper). Water molecules forming a single-layer structure exhibit significantly higher thermal conductivity than the other structures (shown in Fig.~\ref{fig:fig3}a). The thermal conductivity for bulk systems measured in experiments and numerical simulations using the SPC/E water model at similar temperatures is about $0.6-0.85 \, \mathrm{W/m\: K}$ \cite{huber2012new, romer2012nonequilibrium, alkhwaji2021selected}. The thermal conductivity can be at least $10 \, \mathrm{\%}$ higher than these values when the water molecules are in a single-layer structure and can be lower in other structures conversely. Moreover, comparing $\kappa$ especially in the single-file structure, it tends to be higher at a smaller radius of CNT.

We further investigate the effects of the structure and behavior of water molecules in the CNT on thermal conductivity. Previous simulations have reported that water confined in nanotubes has better transport properties, including heat conduction, in the longitudinal direction than in the radial direction \cite{liu2005transport, liu2005dynamics}. However, the details of the relationship between heat conduction and the motion of molecules perpendicular to it have not been clarified.  {Figures~\ref{fig:fig3}b and \ref{fig:fig3}c indicate the differences in the number density distribution between single-file structure and single-layer structure. In the case of $R=0.352 \, \mathrm{nm}$ and single layer structure, the peak of the density distribution is higher than that of $R=0.391$ and $0.469 \, \mathrm{nm}$ (Fig.~\ref{fig:fig3}b). This tendency corresponds to its high thermal conductivity. By densely congining of the molecules at the same distance from the center of the CNT, the energy propagation due to the interaction between molecules becomes more linear in the long-axis direction, resulting in higher thermal conductivity. Regarding $R=0.352 \, \mathrm{nm}$, the peak of the density distribution in the radial direction is much higher than that of a single-layer structure, while the density along the long axis direction is far lower (Fig.~\ref{fig:fig3}c). When the density of water molecules is low, the distance between oxygen and hydrogen atoms that form hydrogen bonds increases. That results in weakening their interaction, and the thermal conductivity decreases. See Fig.~S2 for the differences in the density distribution of water molecules between single-layer structure and double-layer structure.

To examine the transition between layers, Figures~\ref{fig:fig3}d and \ref{fig:fig3}e show a variation in the distance of a typical water molecule from the central axis of the CNT. Taking note of where the water molecules are located in the radial direction of the nanotube, there is a crucial difference between the single-layer structure and the double-layer structure. In the former, the distance $r$ from the central axis of the CNT ($r=0$) is nearly constant due to the vibration within one shell, while in the latter, it goes back and forth between two values. This means each molecule transitions to another layer at irregular intervals.  When a water molecule moves to another layer, energy is consumed to break hydrogen bonds and the regular structure is also disrupted, making it difficult for energy to conduct in the longitudinal direction.

\section{Conclusion}
We have conducted non-equilibrium MD simulations to measure the heat conduction of water molecules inside the CNT. In the case where the radius is very small, water molecules have a pronounced confinement effect and align in a single file within the CNT. This alignment leads to a divergent thermal conductivity, based on KPZ class: $\kappa \sim L^{\alpha}$. The power-law exponent $\alpha =0.33$ obtained here was also experimentally observed in other one-dimensional heat transport systems \cite{hsiao2013observation,tang2016length,lee2017divergent}. As the CNT radius increases, water molecules form a single-layer structure and a double-layer structure due to weakened confinement. Consequently, the thermal conductivity of water molecules becomes a constant value, regardless of the CNT length, similar to the behavior observed in bulk water. The crossover from 1D to 3D liquid state, evident when increasing the radius, has been also observed in simulations using multiparticle collision dynamics \cite{luo2021heat}. This implies that dimension-dependent thermal properties of the liquid can be controlled. On the other hand, the length-dependent heat conduction of solid CNTs shows almost the equivalent power-law divergence regardless of the radius \cite{cao2012size}.

Our research indicates that the fluidity of molecules plays an influential role in heat transport, suggesting a distinct mechanism from the phonon scattering-driven heat transport observed in solid systems. Remarkably, we further provide a novel relationship between thermal conductivity and the structure of water molecules.  The thermal conductivity of water is higher in molecules with a single-layer structure compared to those forming alternative configurations. This is caused by configuration and motion in the direction perpendicular to the heat flux, such as molecular density distribution in the radial direction and transitions to another layer. Thus, transferring energy more directly in the direction of heat flux and reducing energy consumption not in the direction of the heat flux would be the key to increasing thermal conductivity. In addition, the thermal conductivity in the range of $0.8$ nm $<R<$ 0.9 nm, transit from double layer to triple layer, is also comparable to the thermal conductivity in 0.6 nm $<R<$ 0.8 nm (see Fig. 3). Therefore, when $R$ attains a sufficient size, thermal conductivity is considered to be independent of $R$, emphasizing the outstanding thermal conductivity of single-layer structural arrangements. 
%\textcolor{red}{In our systems, pressures of 1.0 MPa are applied to the water molecules in the reservoirs by the pistons. Water in a bulk system has little pressure dependence on thermal conductivity unless very high pressures are applied \cite{ross1984thermal}. However, in our system, the piston pressure changes the density of water molecules in the CNTs, which may affect the thermal conductivity. However, in our present results, even if the value of thermal conductivity changes, it does not seem to affect the trend. That is because pressure is not related to the qualitative result that the structure of water molecules due to confinement affects thermal conductivity.}

The relationship between molecular behavior and thermal conductivity, as dependent on the radius of the CNT revealed in this study, will aid in understanding the principles of thermal transport at a microscopic level. Moreover, the prospect of enhancing heat conduction even with the same material by mere adjustments in the system's length and width could pave a pathway for the development of heat transport devices in engineering applications.

%\bibliography{Thesis1}
%\bibliographystyle{nature}

%merlin.mbs apsrev4-1.bst 2010-07-25 4.21a (PWD, AO, DPC) hacked
%Control: key (0)
%Control: author (8) initials jnrlst
%Control: editor formatted (1) identically to author
%Control: production of article title (-1) disabled
%Control: page (0) single
%Control: year (1) truncated
%Control: production of eprint (0) enabled
%

% \bibliographystyle{ScienceAdvances}
% \bibliography{/Users/yamamoto/Dropbox/bibsource/crowding,/Users/yamamoto/Dropbox/bibsource/crowding_phase_bio,/Users/yamamoto/Dropbox/bibsource/diffusion,/Users/yamamoto/Dropbox/bibsource/diffusion_hetero,/Users/yamamoto/Dropbox/bibsource/diffusion_reaction,/Users/yamamoto/Dropbox/bibsource/force_field,/Users/yamamoto/Dropbox/bibsource/force_field_CG,/Users/yamamoto/Dropbox/bibsource/MD_method,/Users/yamamoto/Dropbox/bibsource/My_Collection,/Users/yamamoto/Dropbox/bibsource/noise,/Users/yamamoto/Dropbox/bibsource/power_law,/Users/yamamoto/Dropbox/bibsource/protein_folding,/Users/yamamoto/Dropbox/bibsource/protein_IDP,/Users/yamamoto/Dropbox/bibsource/protein_loop,/Users/yamamoto/Dropbox/bibsource/protein_stability,/Users/yamamoto/Dropbox/bibsource/protein_villin_Fip35,/Users/yamamoto/Dropbox/bibsource/search,/Users/yamamoto/Dropbox/bibsource/water_around_protein,/Users/yamamoto/Dropbox/bibsource/polymer}

\end{document}